\documentclass[10pt,twocolumn,letterpaper]{article}

\usepackage[pagenumbers]{cvpr} 

\usepackage{graphicx}
\usepackage{amsmath}
\usepackage{amssymb}
\usepackage{booktabs}
\usepackage{float}
\usepackage{placeins}
\usepackage{stfloats}
\usepackage{amsmath}
\usepackage{xcolor}
\usepackage{enumitem}

\usepackage[pagebackref,breaklinks,colorlinks]{hyperref}

\usepackage[capitalize]{cleveref}
\crefname{section}{Sec.}{Secs.}
\Crefname{section}{Section}{Sections}
\Crefname{table}{Table}{Tables}
\crefname{table}{Tab.}{Tabs.}

\newcommand{\visLabel}[1]{{\fontfamily{cmtt}\selectfont {#1}}}
\definecolor{foo}{HTML}{EEEEFF}

\def\cvprPaperID{*****} 
\def\confName{URBANACCESS'22}
\def\confYear{2022}

\begin{document}

\title{Crowdsourcing and Sidewalk Data:\\ A Preliminary Study on the Trustworthiness of OpenStreetMap Data in the US}

\author{
\textbf{Kazi Shahrukh Omar$^{1}$, Gustavo Moreira$^{1}$, Daniel Hodczak$^{1}$, }\\
\textbf{Maryam Hosseini$^{2}$, Fabio Miranda$^{1}$}\\
\fontsize{10pt}{10pt}\selectfont $^{1}$~University of Illinois Chicago, $^{2}$~New York University
}

\twocolumn[{%
\renewcommand\twocolumn[1][]{#1}%
\maketitle
\begin{center}
    \centering
    \captionsetup{type=figure}
    \includegraphics[width=\linewidth]{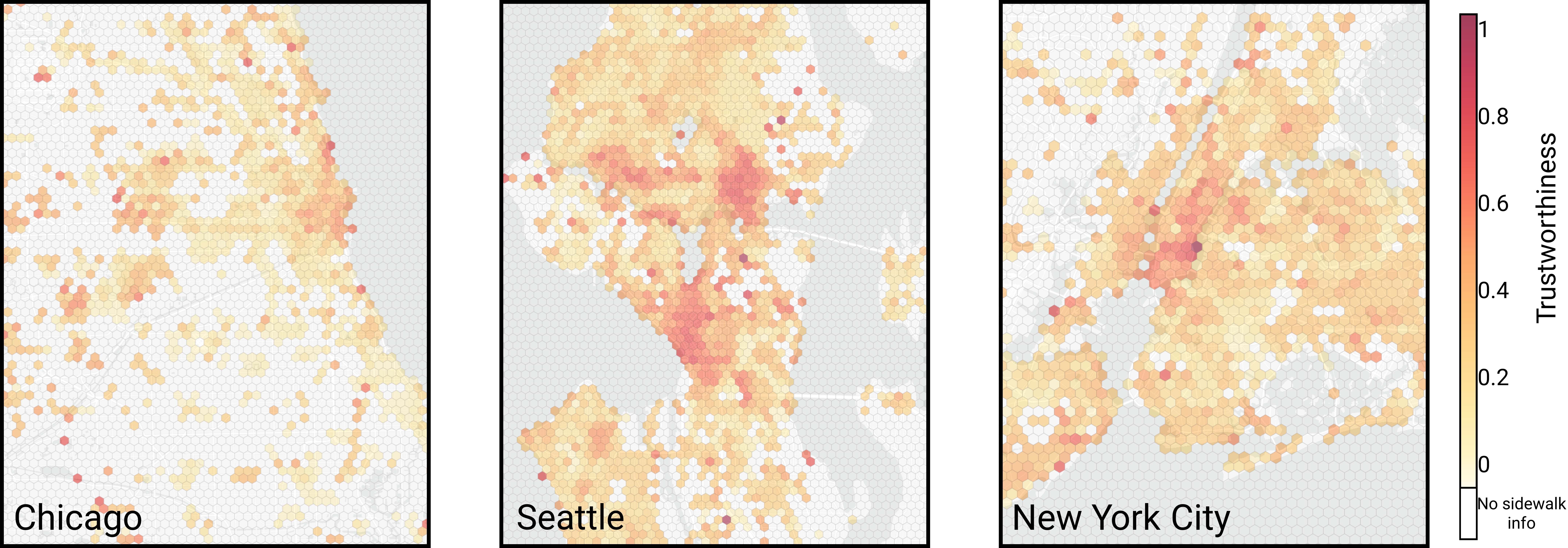}
    \captionof{figure}{Spatial distribution of the average trustworthiness of \emph{roads with sidewalk information} and \emph{sidewalk geometries} in Chicago, Seattle and New York City.}
    \label{fig:hexbin_trust}
\end{center}%
}]

\begin{abstract}
Sidewalks play a pivotal role in urban mobility of everyday life. Ideally, sidewalks provide a safe walkway for pedestrians, link public transportation facilities, and equip people with routing and navigation services. 
However, there is a scarcity of open sidewalk data, which not only impacts the accessibility and walkability of cities but also limits policymakers in generating insightful measures to improve the current state of pedestrian facilities. As one of the most famous crowdsourced data repositories, OpenStreetMap (OSM) could aid the lack of open sidewalk data to a large extent. However, completeness and quality of OSM data have long been a major issue. In this paper, we offer a preliminary study on the availability and trustworthiness of OSM sidewalk data. First, we compare OSM sidewalk data coverage in over 50 major cities in the United States. Then, we select three major cities (Seattle, Chicago, and New York City) to further analyze the completeness of sidewalk data and its features, and to compute a trustworthiness index leveraging historical OSM sidewalk data.
\end{abstract}


\begin{figure*}[t]
  \centering
  \includegraphics[width=\linewidth]{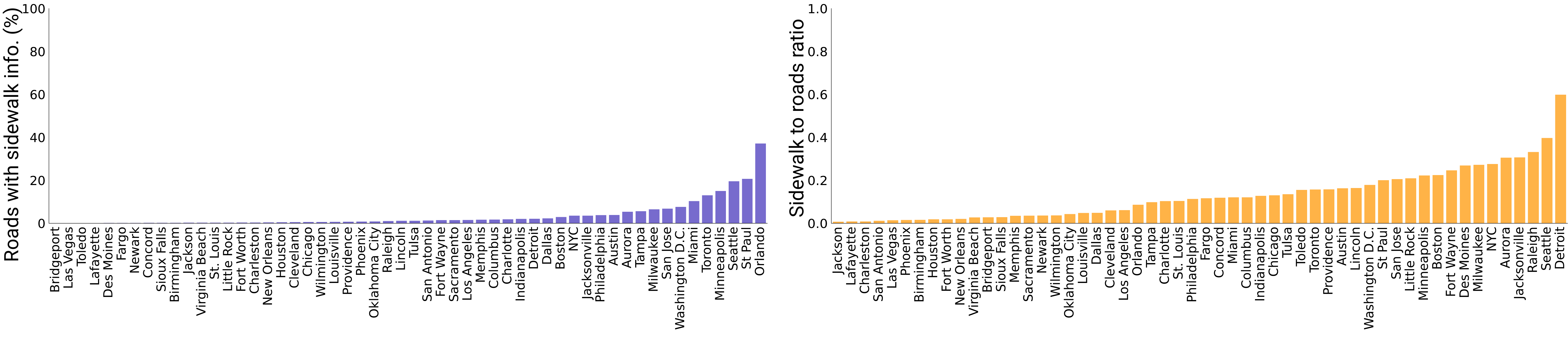}
  \vspace{-0.5cm}
  \caption{Comparing the coverage of sidewalk data in OSM for 54 cities in the US. Left: percentage of roads with sidewalk information. Right: ratio between sidewalks and roads. The majority of cities have less than 20\% of roads with sidewalk information. In addition, only Seattle and Detroit have a sidewalk to road ratio above 0.4.}
  \label{fig:sidewalk_coverage}
  \vspace{-0.5cm}
\end{figure*}

\section{Introduction}

Sidewalks are arguably the most important pedestrian-dedicated public spaces. They are the focal point of cities at the human scale, where the most widely used and environmentally sustainable form of transportation (walking/rolling) takes place~\cite{vale2016active, twardzik2019features}. Sidewalks can significantly impact the everyday life of people, specifically for those relying on such infrastructure as the primary means of accessing public spaces~\cite{qin2018pedestrian, saha2019project, ferreira2007proposal, clarke2008mobility, rosenberg2013outdoor} and public transit~\cite{woldeamanuel2016measuring}. Well-designed sidewalks can also support local businesses and promote economic activities~\cite{kim2012mixed, liu2015sidewalk}. %
Despite their prominent role in various pedestrian-level analysis, the available public data on sidewalks is significantly scarce and limited in time and geographical coverage~\cite{deitz2021squeaky}. Ironically, data collection and monitoring of motorized travel patterns and infrastructure have been practiced since the 1950s~\cite{nordback2016exploring}, and the collected data has been extensively used in transportation planning research, while collecting pedestrian-level data has only recently received some attention~\cite{louch2020availability}. 
Furthermore, the existing datasets are collected mainly by resourceful cities, with substantial variation in the extent of data and attributes, and inconsistent methods from place to place. Such problems create significant barriers to conducting comparative studies, or data integration, across administrative borders~\cite{deitz2021free, louch2020availability, HOU2020102772}. Majority of the sidewalk inventories still lack important feature such as width, surface materials, curb ramps, tactile surface indicators, audio signals, etc., which are essential for specific mobility needs~\cite{HOU2020102772}. 
This lack of sidewalk data not only affects accessible routing in cities but also impacts how urban planners and policymakers could scrutinize mobility issues and the condition of sidewalks to make informed decisions~\cite{chin2008accessibility}. 
%
Ongoing projects~\cite{saha2019project, opensidewalks, bolten2019accessmap, hosseini2022citysurfaces, mobasheri2018enrichment, olivatto2019drone, vestena2022osm} propose to standardize and improve sidewalk data coverage. Even so, researchers primarily rely on crowdsourced OpenStreetMap (OSM) data, the largest geospatial open-data initiative, with data covering not only streets and roads, but also buildings, points of interest, and other geographic entities\cite{haklay2008openstreetmap}.

Considering OSM's importance in sidewalk studies, our primary goal is to assess the coverage and trustworthiness of OSM sidewalk data. While previous studies~\cite{mobasheri2015completeness, mobasheri2017crowdsourced} have assessed completeness of sidewalks, we propose to also assess the trustworthiness of the data by analyzing its history and provenance. Concretely, we compare the sidewalk coverage of OSM in 54 major cities in the United States and further expand our analysis in three major cities (Seattle, Chicago and New York City) with the spatial trustworthiness of the data. We then highlight possible research directions to mitigate some of the identified problems.

\section{Related Work}
\label{sec:relatedWork}

OSM data has been used in various studies spanning different fields, including routing \cite{luxen2011real, felicio2022handling}, location-based services \cite{ciepluch2009using, das2014location1, das2014location2}, traffic and transportation \cite{zilske2015openstreetmap, rieck2015advanced, narboneta2016opentripplanner, xu2016cross}, energy modeling \cite{alhamwi2017openstreetmap, schiefelbein2019automated}, population estimation \cite{bakillah2014fine, rosina2017using}, 3D city modeling \cite{over2010generating, Doraiswamy:2018:IVE:3183713.3193559, 8283638}, land cover use \cite{fonte2017generating, fonte2019using}, and emergency response management \cite{ eckle2015quality, poiani2016potential}. 
Ongoing initiatives to improve sidewalk data, including Accessmap \cite{bolten2019accessmap} and OpenSidewalks \cite{opensidewalks}, also utilize data from OSM. 
However, quality of OSM data has always been a major concern for both research and industrial purposes \cite{hecht2013measuring, siebritz2014assessing, husen2018quality, brovelli2018new, basaraner2020geometric, hoong2020assessing, girindran2020reliable, borkowska2022analysis, wu2021comprehensive}.
To tackle this problem, studies have been conducted to evaluate the quality and completeness of OSM data, focusing on different entities, such as roads \cite{jilani2013multi, jilani2014automated, minaei2020evolution}, buildings \cite{brovelli2018new, basaraner2020geometric}, and points of interest \cite{steiniger2016can}.
Properly assessing the quality of OSM data is a challenge, given that traditional approaches rely on the availability of official data (though even such data might have problems, including slow update rate). Alternative approaches then analyze the \emph{evolution} of the data itself, assessing how the number of users editing, confirmations from different users, number of versions and rollbacks contribute to the quality of OSM data \cite{mooney2012social, napolitano2012mvp, kessler2011tracking, kessler2013trust, d2014vgi, zhou2016version, fogliaroni2018data, alghanim2021leveraging}.
Considering this, we propose to evaluate the use of trustworthiness as an index for OSM sidewalk data.


\begin{figure}[b!]
  \centering
  \includegraphics[width=\linewidth]{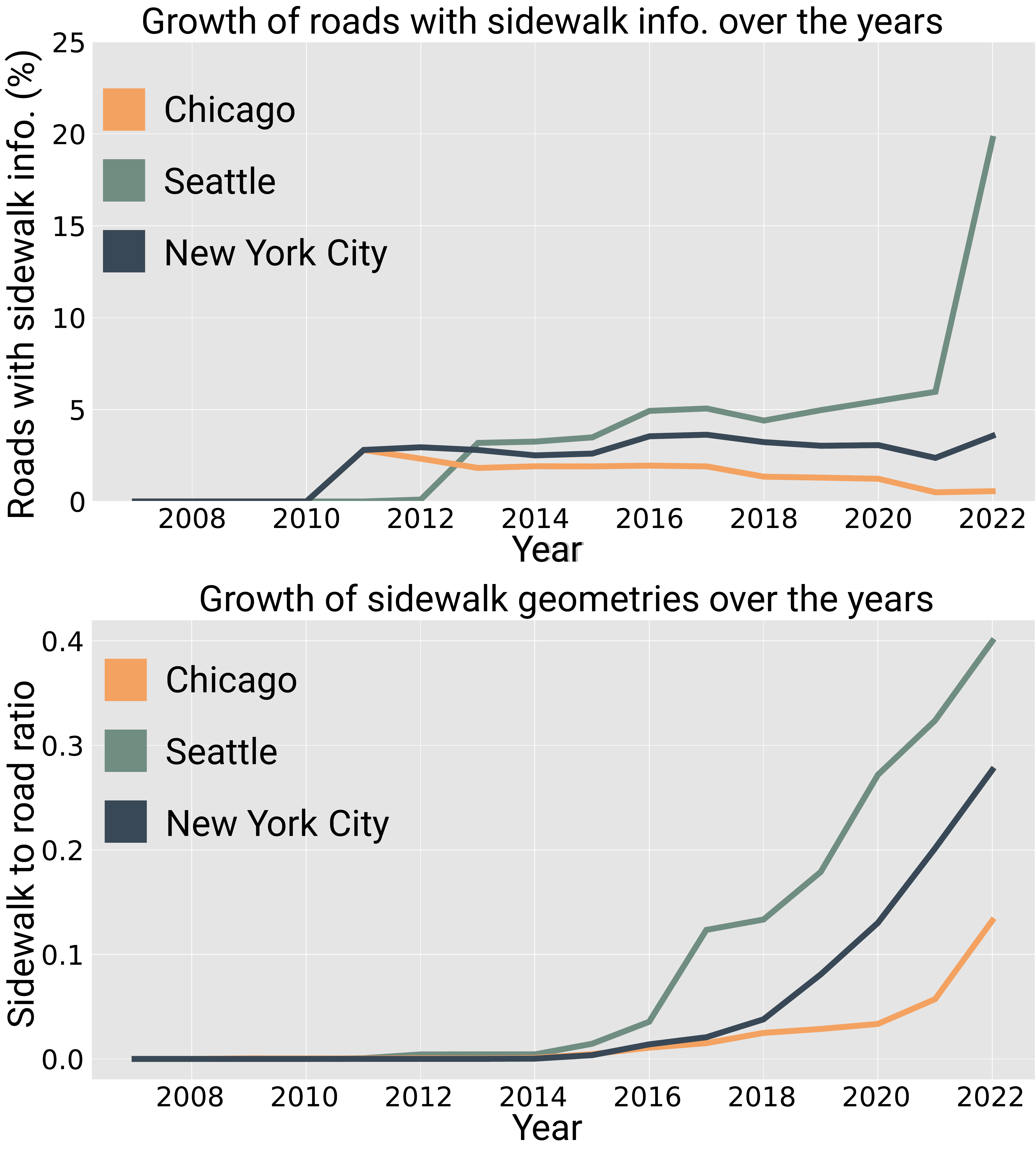}
  \caption{Growth of OSM sidewalk data in Chicago, Seattle, and NYC over the years. Top: growth of roads with sidewalk information. Bottom: growth of sidewalk / road ratio.}
  \label{fig:growth}
\end{figure}

\section{Study Area and OpenStreetMap Data}

Between 2001 and 2019, the US built-up land - buildings, roads, and other infrastructures - increased by more than 14,000 square miles of new developments \cite{urbanGrowth}. Today, more than 80\% of the US population (around 309 million people) lives in urban areas \cite{citiesUS}. Subsequently, there has been a growth in the availability of OSM data across the US, capturing its complex and diverse built environment. 
Our study first analyzes the availability of OSM sidewalk data~(Sec.~\ref{sec:availability}), followed by an analysis on the trustworthiness of the data~(Sec.~\ref{sec:trust}). First, we begin with an overview of the structure of the OSM data and how we use it.

The OSM data model is built upon three basic elements: \emph{nodes} (points in space), \emph{ways} (linear features and boundaries), and \emph{relations} (multi-purpose entities that define relationship between two or more elements -- nodes, ways, and/or other relations).
OSM elements have \textit{tags} attached to them that describe different physical features on the ground (e.g., roads, buildings).
For example, \emph{ways} that have the \emph{highway} tag are used to define highways, roads, streets, or paths. In this paper, highway and road are used interchangeably.
In our analyses, we focus on sidewalks and roads with sidewalk info. In OSM, these features can appear as:

\begin{enumerate}[nosep, leftmargin=*]
    \item Sidewalk as a refinement to a highway: The sidewalk is a property of an existing road. Elements with
    \fcolorbox{white}{foo}{\visLabel{highway=*}}+\fcolorbox{white}{foo}{\visLabel{sidewalk=[both,right,left,yes]}}.
    Here, we excluded OSM road types that cannot have sidewalks along them (i.e., \visLabel{footway, escape, raceway, busway, bridleway, path, cycleway, construction, corridor}).
    \item Sidewalk as a separate \emph{way}: The sidewalk is described as its own separate geometry. Elements with \fcolorbox{white}{foo}{\visLabel{highway=footway}}+\fcolorbox{white}{foo}{\visLabel{footway=sidewalk}}.
\end{enumerate}

In this paper, we make a distinction between the two types, given that sidewalk geometries offer a more spatially accurate representation of the pedestrian environment~\cite{sidewalk-sdss}.
Figure~\ref{fig:sidewalk_coverage} presents the coverage of sidewalk data in OSM for 54 cities, considering both sidewalks as a refinement to a highway and sidewalks as separate ways (i.e., distinct geometry).
Even though there has been a growth in the availability of OSM sidewalk data over the years (Figure \ref{fig:growth}), there is still a lack of data regarding this pedestrian infrastructure.

\begin{figure*}[t]
  \centering
  \includegraphics[width=\linewidth]{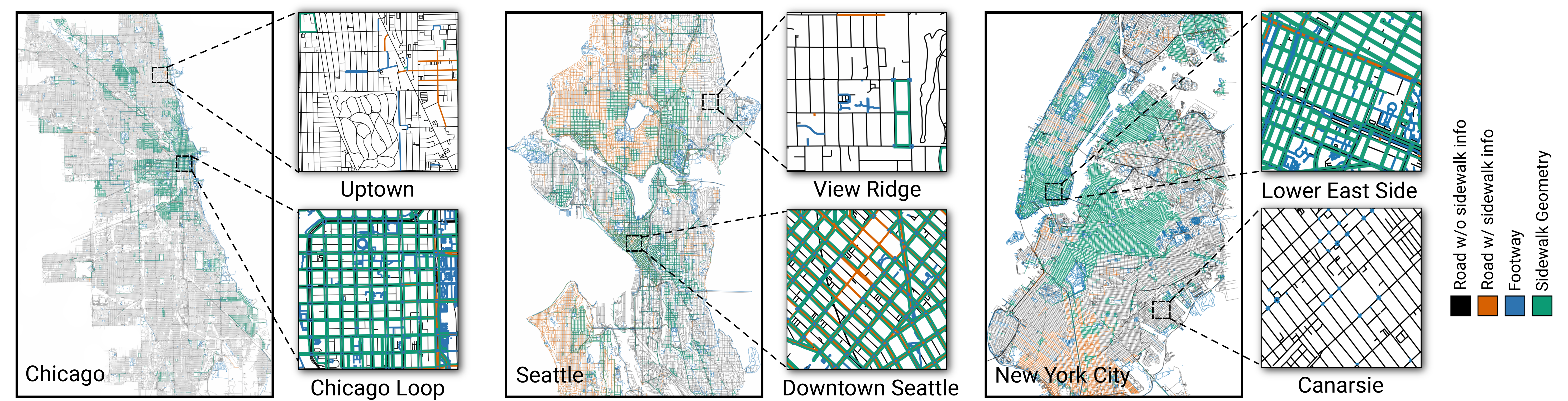}
 \caption{Coverage of \textit{sidewalks} and \textit{roads with sidewalk info} in Chicago, Seattle, and New York City (from left to right). Seattle OSM has comparably higher percentage of sidewalk geometries and roads with sidewalk information mapped than the other two cities. Also, Neighborhoods like \textit{Chicago Loop}, \textit{Downtown Seattle}, and Manhattan's~\textit{Lower East Side} have good sidewalk coverage. On the other hand \textit{Uptown}, \textit{View Ridge}, and \textit{Canarsie} neighborhoods only have a handful of mapped sidewalk geometries.}
  \label{fig:city_visual}
  \vspace{-0.5cm}
\end{figure*}

\begin{figure}[b]
  \centering
  \includegraphics[width=1\linewidth]{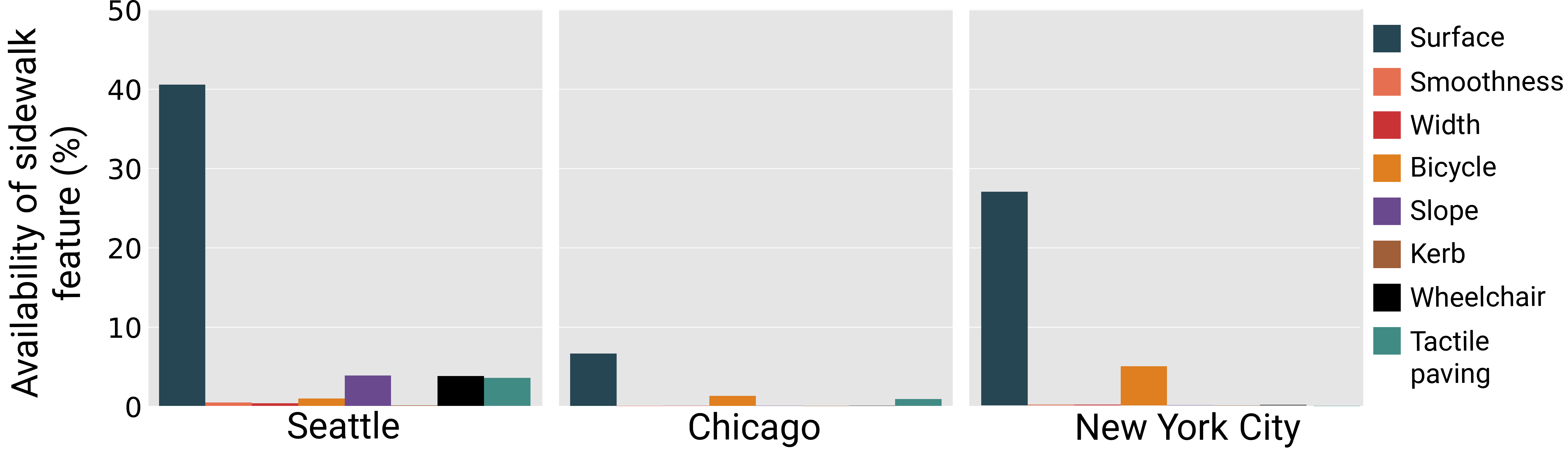}
  \vspace{-0.5cm}
  \caption{Availability of OSM sidewalk attributes. \textit{Surface material} is the most commonly available feature in all three cities whereas kerb, smoothness and width are almost non-existent.}
  \label{fig:sidewalk_tag_availibility}
\end{figure}


\section{Coverage of OSM Sidewalk Data}
\label{sec:availability}

In our work, we measure the object and attribute coverage of OSM sidewalk data. For the object-level analysis, we evaluated the spatial coverage of the following sidewalk categories: roads without sidewalk information, roads with sidewalk information, footway and sidewalk geometry.
Figure~\ref{fig:city_visual} shows the coverage of each category in Chicago, Seattle and New York City.
We can notice how Seattle's data surpasses the other two cities in terms of roads with sidewalk info (19.56\%), and sidewalk to road ratio (0.4). 
Table \ref{table:stat} illustrates the statistics of OSM sidewalk data for all three cities, with Chicago and NYC both having a very low percentage of roads with sidewalk information (respectively 0.55\% and 3.50\%).
Also noteworthy is that there are several regions in these cities with almost no sidewalk data (e.g., Uptown, View Ridge, and Canarsie).

\begin{table*}[t!]
\centering
\caption{Coverage of OSM sidewalk data for selected cities.}
\vspace{-0.25cm}
\begin{tabular}{|l|c|c|c|c|c|}
City & \# of roads & \begin{tabular}[c]{@{}c@{}}\# of sidewalk\\ geometries\end{tabular} & \begin{tabular}[c]{@{}c@{}}\# of roads\\ with sidewalk info\end{tabular}  & \begin{tabular}[c]{@{}c@{}}Percentage of roads\\ with sidewalk info (\%)\end{tabular} & Sidewalk to road ratio\\
\hline
Chicago & 134,598 & 17,492 & 746 & 0.55 & 0.12\\
Seattle & 44,013 & 17,487 & 8,610 & 19.56 & 0.40\\
New York City     & 180,255 & 49,736 & 6,311 & 3.50 & 0.28\\
\end{tabular}
\label{table:stat}
\vspace{-0.5cm}
\end{table*}

For the attribute-level analysis, we assessed the number of additional attributes available in OSM sidewalk geometries (e.g., surface material, width, wheelchair accessibility). These features are of great importance in routing and navigation for people with mobility or vision impairments. Figure~\ref{fig:sidewalk_tag_availibility} shows the availability of different features in the chosen cities. The most commonly present feature in all three cities (surface material) varied between 40.5\% in Seattle, 27\% in NYC, and less than 10\% in Chicago. Some of the attributes (such as kerb, smoothness, and width) were almost non-existent in all three cities.

\section{Trustworthiness of OSM Sidewalk Data}
\label{sec:trust}

\begin{figure}[h]
  \centering
  \includegraphics[width=\linewidth]{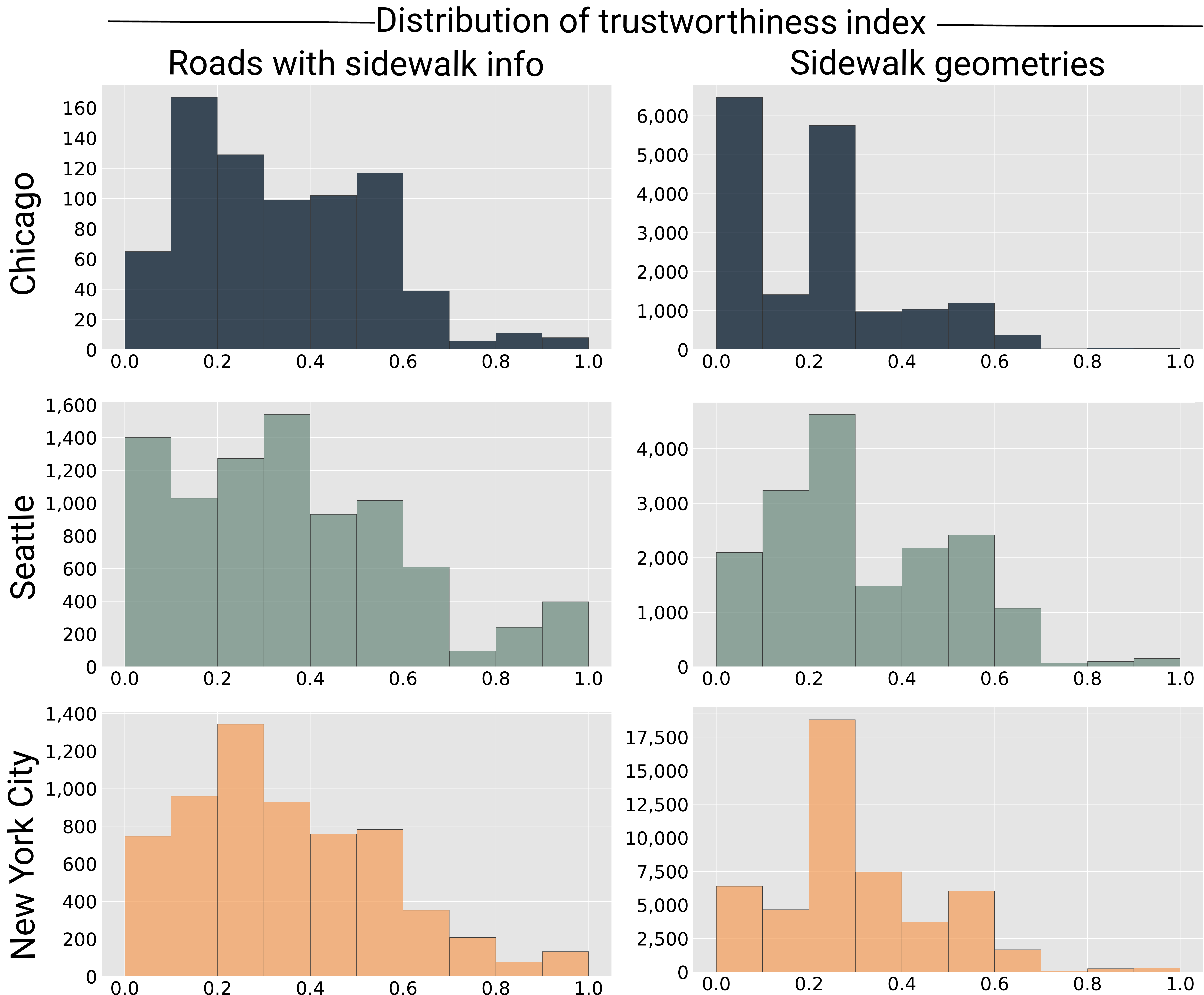}
  \vspace{-0.5cm}
  \caption{Distribution of trustworthiness index of \textit{roads with sidewalk information} and \textit{sidewalk geometries} in the selected cities.}
  \label{fig:trustworthiness}
  \vspace{-0.75cm}
\end{figure}

An important challenge in assessing the quality of OSM data is the lack of official data. Previous studies have then proposed to analyze the evolution of the data itself (history and provenance)~\cite{d2014vgi,fogliaroni2018data, alghanim2021leveraging}, evaluating edit history \cite{mooney2012social}, number of editors \cite{napolitano2012mvp}, number of rollbacks and direct confirmations from different users \cite{kessler2011tracking}.
In our work, we make use of Alghanim et al.'s approach \cite{alghanim2021leveraging} to compute a trustworthiness index of OSM sidewalk data, based on direct, indirect and temporal components. Following their approach, trustworthiness is defined as follows:

\vspace{-0.5cm}
\begin{equation*}
T(f_i) = w_{d} * T_{dir}(f_i) + w_{i} * T_{ind}(f_i) + w_{time} * T_{time}(f_i)
\label{eq-1}
\end{equation*}

\noindent where $T_{dir}(f_i)$, $T_{ind}(f_i)$ and $T_{time}(f_i)$ are the direct, indirect and temporal indicators of the $i-th$ version of feature $f$, and $w_{dir}$, $w_{ind}$, and $w_{time}$ are weights balancing the components (0.5, 0.25, 0.25, respectively).
A direct indicator depends on the feature version historical information.
An indirect indicator models contributions that do not directly depend on the current feature version, but its neighboring spatial features.
A temporal indicator models the impact of time on a feature trustworthiness, i.e., the longer a feature persists, the higher the chances it actually matches the real feature.
For a more detailed discussion, we refer the reader to Alghanim et al.~\cite{alghanim2021leveraging}.
Figure~\ref{fig:trustworthiness} shows the trustworthiness index distribution for both \emph{roads with sidewalk information} and \emph{sidewalk geometries}, and Tables~\ref{table:dist1} and \ref{table:dist2} highlight the percentage with \emph{high} trustworthiness. Given these preliminary results, not only sidewalk coverage might be an issue, but also the trustworthiness of this data.

\begin{table}[h]
\centering
\caption{Trustworthiness index of \textit{roads with sidewalk info}.}
\vspace{-0.25cm}
\begin{tabular}{c|c|c}
City              & T-index $<$ 0.5 (\%)    & T-index $\geq$ 0.5 (\%)     \\ \hline
Chicago           & 75.64             & 24.36                  \\ \hline
Seattle           & 72.31             & 27.69                \\ \hline
NYC               & 75.27             & 24.73                    \\
\end{tabular}
\label{table:dist1}
\vspace{-0.5cm}
\end{table}

\begin{table}[h]
\centering
\caption{Trustworthiness index of \textit{sidewalk geometries}.}
\vspace{-0.25cm}
\begin{tabular}{c|c|c}
City              & T-index $<$ 0.5 (\%)    & T-index $\geq$ 0.5 (\%)     \\ \hline
Chicago           & 90.23            & 9.77                  \\ \hline
Seattle           & 78.09            & 21.91                \\ \hline
NYC               & 83.0             & 17.0                    \\
\end{tabular}
\label{table:dist2}
\end{table}

\section{Discussion and Future Work}

In this study, we take steps towards understanding the quality OSM sidewalk data and evaluating both its coverage and trustworthiness, going beyond previous studies that only focused on coverage. The significance of sidewalks to everyday life of urban dwellers coupled with the lack of publicly available datasets describing their locations and attributes, underscore the importance of standardized datasets such as OSM, specifically in the absence of authoritative data. Although preliminary, our goal is to shed light on the availability of OSM sidewalk data, as well providing a method to analyze whether and to what extent the available OSM sidewalk data can be trusted. 
For example, our results show that even among the major and often resourceful cities in the US, there is a significant lack of OSM sidewalk data (Figure~\ref{fig:sidewalk_coverage}). Among the 54 cities analyzed, only 11\% have more than 10\% of their roads tagged with sidewalk information, and 80\% of cities fall below 5\%. In Orlando, ranked first, only 37\% of the roads have any sidewalk-related tags. 

Regarding attributes in the three selected cities, surface material has the highest availability,  with 40.4\% of Seattle's sidewalks, 26.9\% of NYC's sidewalks, and 6.6\% of Chicago's sidewalks having this information.  
The remaining attributes were found in insignificant quantities or were absent in the select cities. This analysis further illustrates the inadequacy of the existing data in supporting comprehensive analysis of sidewalks~\cite{hosseini2022towards}. 
In future work, we plan to further evaluate the measures adopted in this study, using authoritative data to validate the trustworthiness and correctness of the data. We will also investigate new techniques to extract important sidewalk attributes.

{\small
\bibliographystyle{unsrt}
\bibliography{paper}
}

\end{document}